\documentstyle[sprocl]{article}
\input{psfig}
\def\be{\begin{equation}}
\def\ee{\end{equation}}
\def\bea{\begin{eqnarray}}
\def\eea{\end{eqnarray}}

\begin{document}
\thispagestyle{empty}
\title{
{\small\rm\vspace*{-3.2cm}
\rightline{TTP97-21}
\rightline{hep-ph/9705404}
\rightline{May 1997}
\ \\
\ \\}
CP VIOLATION IN $B_s$ DECAYS 
\footnote{Invited plenary talk given at the {\it 2nd International 
Conference on B Physics and CP Violation}, Honolulu, Hawaii, March 24--27, 
1997. To appear in the proceedings.}
}
\author{ROBERT FLEISCHER}
\address{Institut f\"ur Theoretische Teilchenphysik, Universit\"at
Karlsruhe, \mbox{D--76128 Karlsruhe}, Germany}
\maketitle\abstracts{CP-violating effects in non-leptonic $B_s$-meson decays 
are reviewed. Special emphasis is given to implications arising from the 
width difference $\Delta\Gamma_s$ between the $B_s$ mass eigenstates. 
If $\Delta\Gamma_s$ is found to be sizable, certain untagged $B_s$-meson 
decays may allow interesting probes of CP violation, extractions of the 
CKM angle $\gamma$ and the Wolfenstein parameter $\eta$, and may 
indicate physics beyond the Standard Model.}
\section{Introduction}\label{intro}
A characteristic feature of the neutral $B_q$-meson systems $(q\in\{d,s\})$
is $B^0_q-\overline{B^0_q}$ mixing. The corresponding time-evolutions are
governed by the $B_q$ mass eigenstates $B_q^{\mbox{{\scriptsize Heavy}}}$ 
and $B_q^{\mbox{{\scriptsize Light}}}$ which are characterized by their mass 
eigenvalues $M_H^{(q)}$, $M_L^{(q)}$ and decay widths $\Gamma_H^{(q)}$, 
$\Gamma_L^{(q)}$. Because of these mixing effects, oscillatory $\Delta M_q t$ 
terms with $\Delta M_q\equiv M_H^{(q)}-M_L^{(q)}$ show up in the 
time-dependent transition rates $\Gamma(B^0_q(t)\to f)$ and 
$\Gamma(\overline{B^0_q}(t)\to f)$ describing decays of initially present 
$B^0_q$ and $\overline{B^0_q}$ mesons into a final state $f$, respectively. 
Essentially all the information that is needed to evaluate these decay rates 
is contained in the convention-independent observable \cite{F97}
\begin{equation}\label{e2}
\xi_f^{(q)}=\exp\left[-i\,\Theta_{M_{12}}^{(q)}\right]
\frac{A(\overline{B^0_q}\to f)}{A(B^0_q\to f)}\,,
\end{equation}
where $\Theta_{M_{12}}^{(q)}$ is the weak $B_q^0$--$\overline{B_q^0}$ mixing
phase that is related to 2\,arg$(V_{tq}^\ast V_{tb})$ and the $A$'s denote
the decay amplitudes corresponding to $\overline{B^0_q}\to f$ and 
$B^0_q\to f$.

It is a well-known feature that $\xi_f^{(q)}$ can be calculated in a
clean way if $B_q\to f$ is dominated by a single CKM amplitude (see 
e.g.\ Ref.~\cite{F97} for a recent discussion). The ``gold-plated'' mode in
this respect is $B_d\to J/\psi\, K_{\rm S}$ measuring $\sin(2\,\beta)$ with
excellent accuracy through mixing-induced CP violation.\cite{desh}\, The 
$B_s$ system plays an important role to determine $\gamma$ -- another angle
of the usual ``non-squashed'' unitarity triangle of the CKM matrix. 

A decay that was proposed frequently in the previous literature to accomplish 
this task is $B_s\to\rho^0K_{\rm S}$. However, this mode is unfortunately
not dominated by a single CKM amplitude. Here penguin contributions lead 
to serious problems so that
\begin{equation}
\xi^{(s)}_{\rho^0K_{\rm S}}\approx\exp(-i\,2\,\gamma)
\end{equation}
is a very bad approximation.\cite{F97}\, Consequently $B_s\to\rho^0K_{\rm S}$
should be the ``wrong'' way to extract $\gamma$.
Needless to note, the branching ratio of that decay is expected to be of 
${\cal O}(10^{-7})$ which makes its experimental investigation very 
difficult. Before focussing on other $B_s$ modes that do allow 
meaningful determinations of $\gamma$, let us turn to an experimental 
problem of $B_s$ decays that is related to time-dependent measurements.
\section{The $B_s$ System in Light of $\Delta\Gamma_s$}\label{DeltaG}
The ``strength'' of the $B^0_q-\overline{B^0_q}$ oscillations is measured
by the mixing parameter $x_q\equiv\Delta M_q/\Gamma_q$, where
$\Gamma_q\equiv(\Gamma^{(q)}_H+\Gamma^{(q)}_L)/2$. While we have
$x_d=0.72\pm0.03$ in the $B_d$ system, a large $B_s$ mixing parameter 
$x_s={\cal O}(20)$ is expected within the Standard Model implying very 
rapid $B^0_s-\overline{B^0_s}$ oscillations (see e.g.\ Ref.~\cite{bf-rev}). 
In order to keep track of the corresponding $\Delta M_s t$ terms in the 
time-dependent $B_s$ decay rates, an excellent vertex resolution system is 
required which is a formidable experimental task.

It may, however, not be necessary to trace these rapid $\Delta M_s t$ 
oscillations in order to obtain insights into the mechanism of CP 
violation.\cite{dunietz}\, This remarkable feature is due to the expected 
sizable width difference $\Delta\Gamma_s\equiv\Gamma_H^{(s)}-
\Gamma_L^{(s)}$ originating mainly from CKM favored $\bar b\to \bar cc\bar s$ 
transitions into final states that are common both to $B_s^0$ and 
$\overline{B_s^0}$. The width difference $\Delta\Gamma_s$ may be as large as 
${\cal O}(20\%)$ of the average decay width $\Gamma_s$ as is indicated by
box diagram calculations \cite{boxes}, a second complementary approach 
\cite{excl} where one sums over many exclusive $\bar b\to\bar cc\bar s$
modes, and by an approach using the ``Heavy Quark Expansion'' leading to 
the most recent result \cite{HQE} 
$|\Delta\Gamma_s|/\Gamma_s=0.16^{+0.11}_{-0.09}$. This width difference
can be determined experimentally e.g.\ from angular correlations in $B_s\to
J/\psi\,\phi$ decays.\cite{ddlr}\, One expects $10^3-10^4$ reconstructed
$B_s\to J/\psi\,\phi$ events both at Tevatron Run II and at HERA-B which
may allow a precise measurement of $\Delta\Gamma_s$.

Because of this width difference already untagged $B_s$ rates, which are 
defined by 
\begin{equation}\label{e1}
\Gamma[f(t)]\equiv\Gamma(B_s^0(t)\to f)+\Gamma(\overline{B^0_s}(t)\to f)\,,
\end{equation}
may provide valuable information about the phase structure of the observable
$\xi_f^{(s)}$ defined by Eq.~(\ref{e2}). This can be seen nicely by writing 
Eq.~(\ref{e1}) in a more explicit way as follows:
\begin{equation}\label{e3}
\Gamma[f(t)]\propto\left[\left(1+\left|\xi_f^{(s)}
\right|^2\right)\left(e^{-\Gamma_L^{(s)} t}+e^{-\Gamma_H^{(s)} t}\right)
-2\mbox{\,Re\,}\xi_f^{(s)}\left(e^{-\Gamma_L^{(s)} t}-
e^{-\Gamma_H^{(s)} t}\right)\right].
\end{equation}
In this expression the rapid oscillatory $\Delta M_s t$ terms, which
show up in the tagged rates, cancel. Therefore it depends only on the
two exponents $e^{-\Gamma_L^{(s)} t}$ and $e^{-\Gamma_H^{(s)} t}$. From 
an experimental point of view such untagged analyses are clearly much 
more promising than tagged ones in respect of efficiency, acceptance 
and purity. 

In order to illustrate these untagged strategies in more detail, let me 
discuss an estimate of $\gamma$ from $B_s\to K^+K^-$ and 
$B_s\to K^0\overline{K^0}$ decays.\cite{fd1}\, Using the $SU(2)$ isospin 
symmetry of strong interactions to relate the QCD penguin contributions 
to these decays 
(electroweak penguins are color-suppressed in these modes and thus play a 
minor role \cite{F97}) yields
\begin{equation}\label{e4}
\Gamma[K^+K^-(t)]\propto |P|^2\Bigl[\bigl(1-2\,|r|\cos\rho\,
\cos\gamma+|r|^2\cos^2\gamma\bigr)e^{-\Gamma_L^{(s)} t}+|r|^2\sin^2\gamma\, 
e^{-\Gamma_H^{(s)} t}\Bigr]
\end{equation}
and
\begin{equation}\label{e5}
\Gamma[K^0\overline{K^0}(t)]\propto |P|^2\,e^{-\Gamma_L^{(s)} t}\,,
\end{equation}
where 
\begin{equation}\label{e6}
r\equiv|r|e^{i\rho}=\frac{|T|}{|P|}e^{i(\delta_{T}-\delta_{P})}\,.
\end{equation}
Here $P$ denotes the $\bar b\to\bar s$ QCD penguin amplitude, 
$T$ is the color-allowed $\bar b\to\bar uu\bar s$ tree amplitude, and 
$\delta_{P}$ and $\delta_{T}$ are the corresponding CP-conserving strong 
phases. In order to determine $\gamma$ from the untagged rates 
Eqs.~(\ref{e4}) and (\ref{e5}), an additional input is needed. Using
the $SU(3)$ flavor symmetry of strong interactions to this end and 
neglecting color-suppressed current-current contributions to 
$B^+\to\pi^+\pi^0$ gives
\begin{equation}\label{e7}
|T|\approx\lambda\,\frac{f_K}{f_\pi}\,\sqrt{2}\,|A(B^+\to\pi^+\pi^0)|\,,
\end{equation}
where $\lambda=0.22$ is the Wolfenstein parameter \cite{wolf}, $f_K$ and
$f_\pi$ are the $K$ and $\pi$ meson decay constants, respectively,
and $A(B^+\to\pi^+\pi^0)$ denotes the appropriately normalized 
$B^+\to\pi^+\pi^0$ decay amplitude. Since $|P|$ is known from
$B_s\to K^0\,\overline{K^0}$, the quantity $|r|=|T|/|P|$ can be estimated 
with the help of Eq.~(\ref{e7}) and allows an estimate of $\gamma$
from the part of Eq.~(\ref{e4}) evolving with exponent 
$e^{-\Gamma_H^{(s)} t}$. If more reliable ways to fix $|T|$ should become
available in the future, this ``estimate'' of $\gamma$ may well turn into 
a solid determination. 
\section{$B_s$ Decays into Admixtures of CP Eigenstates}\label{admixtures}
One can even do better than in the previous section, i.e.\ without using 
$SU(3)$ flavor symmetry, by considering decays corresponding 
to $B_s\to K \overline{K}$ where two vector mesons or higher resonances 
are present in the final states.\cite{fd1}\,  
\subsection{An Extraction of $\gamma$ using $B_s\to K^{\ast+}
K^{\ast-}$ and $B_s\to K^{\ast0}\overline{K^{\ast0}}$}\label{kkbar}
The untagged angular distributions of these decays, which are given 
explicitly in Ref.~\cite{fd1}, provide many more observables than the 
untagged modes $B_s\to K^+K^-$ and $B_s\to K^0\overline{K^0}$ discussed in 
the previous section. In the case of $B_s\to K^{\ast0}\overline{K^{\ast0}}$
the formulae simplify considerably since it is a penguin-induced 
$\bar b\to\bar sd\bar d$ mode and receives therefore no tree contributions.
Using again the $SU(2)$ isospin symmetry of strong interactions, the 
QCD penguin contributions to $B_s\to K^{\ast+}K^{\ast-}$ and 
$B_s\to K^{\ast0}\overline{K^{\ast0}}$ can be related to each other. 
If one takes into account these relations and goes very carefully through 
the observables of the untagged angular distributions, one 
finds that they allow the extraction of the CKM angle 
$\gamma$ without any additional theoretical input.\cite{fd1}\, Needless to
note, the angular distributions provide moreover information about the 
hadronization dynamics of these decays. Since the formalism \cite{fd1} for
$B_s\to K^{\ast+}K^{\ast-}$ applies also to $B_s\to\rho^0\phi$, it may allow 
insights into the physics of electroweak penguins as the latter mode is 
dominated by these operators.\cite{F97}\,
\subsection{The ``Gold-plated'' Transitions to Extract $\eta$}\label{gold}
This subsection is devoted to the decays $B_s\to D_s^{\ast+}D_s^{\ast-}$ 
and $B_s\to J/\psi\,\phi$, which is the counterpart of the ``gold-plated'' 
mode $B_d\to J/\psi\,K_{\rm S}$ to measure the CKM angle $\beta$. Since
these decays are dominated by a single CKM amplitude, the hadronic 
uncertainties cancel in $\xi_f^{(s)}$ which takes in that particular case 
the form \cite{fd1}
\begin{equation}\label{e11}
\xi_f^{(s)}=\exp(i\,\phi_{\rm CKM})\,.
\end{equation}
Consequently the observables of the angular distributions simplify 
considerably. A characteristic feature of these angular 
distributions is interference between CP-even and CP-odd final state 
configurations leading to observables that are proportional to  
\begin{equation}\label{e12}
\left(e^{-\Gamma_L^{(s)}t}-e^{-\Gamma_H^{(s)}t}
\right)\sin\phi_{\rm CKM}\,.
\end{equation}
Here the CP-violating weak phase is given by $\phi_{\rm CKM}=
2\lambda^2\eta\approx{\cal O}(0.03)$, where $\lambda$ and $\eta$ 
are two of the Wolfenstein parameters.\cite{wolf}\, The observables of the 
angular distributions for both the color-allowed channel 
$B_s\to D_s^{\ast+} D_s^{\ast-}$ and 
the color-suppressed transition $B_s\to J/\psi\,\phi$ each provide  
sufficient information to determine the CP-violating weak phase 
$\phi_{\rm CKM}$ from their untagged data samples thereby 
fixing the Wolfenstein parameter $\eta$.\cite{fd1}\,  Note that this 
extraction of $\phi_{\rm CKM}$ is not as clean as that of $\beta$ from 
$B_d\to J/\psi\,K_{\rm S}$. This feature is due to the smallness of 
$\phi_{\rm CKM}$ with respect to $\beta$.

Within the Standard Model one expects a very small value of 
$\phi_{\rm CKM}$ and $\Gamma_H^{(s)}<\Gamma_L^{(s)}$. However, that need 
not to be the case in many scenarios for ``New Physics''.\cite{NP}\,
An experimental study of the decays $B_s\to D_s^{\ast+}D_s^{\ast-}$ and 
$B_s\to J/\psi\,\phi$ may shed light on this issue, and an
extracted value of $\phi_{\mbox{{\scriptsize CKM}}}$ much larger 
than ${\cal O}(0.03)$ would indicate physics beyond the Standard Model.
\section{$B_s$ Decays caused by $\bar b\to\bar uc\bar s$ $(b\to c\bar 
u s)$}\label{nonCP}
The $B_s$ decays discussed in this section are pure tree decays, i.e.\
receive no penguin contributions, and probe the CKM angle $\gamma$ in a 
clean way.\cite{gam}\, 
There are by now well-known strategies on the market using the 
time evolutions of such modes, e.g.\ $\stackrel{{\mbox{\tiny 
(---)}}}{B_s}\to\stackrel{{\mbox{\tiny 
(---)}}}{D^0}\phi$~\cite{gam,glgam} and $\stackrel{{\mbox{\tiny 
(---)}}}{B_s}\to D_s^\pm K^\mp$ \cite{adk}, to extract $\gamma$. However, in
these strategies tagging is essential and the rapid $\Delta M_s t$
oscillations have to be resolved which is an experimental challenge.
The question what can be learned from untagged data samples of 
these decays, where the $\Delta M_s t$ terms cancel, was investigated 
by Dunietz.\cite{dunietz}\, In the untagged case the determination of 
$\gamma$ requires additional inputs: a measurement of the untagged $B_s\to 
D^0_{\rm CP} \phi$ rate in the case of the color-suppressed 
modes $\stackrel{{\mbox{\tiny (---)}}}{B_s}\to\stackrel{{\mbox{\tiny 
(---)}}}{D^0}\phi$, and a theoretical input corresponding to the
ratio of the unmixed rates $\Gamma(B^0_s\to D_s^-K^+)/\Gamma(B^0_s\to
D_s^-\pi^+)$ in the case of the color-allowed decays
$\stackrel{{\mbox{\tiny (---)}}}{B_s}\to D_s^\pm K^\mp$. This ratio can 
be estimated with the help of the ``factorization'' hypothesis which may 
work reasonably well for these color-allowed channels.\cite{stech}

Interestingly the untagged data samples may exhibit CP-violating 
effects that are described by observables of the form
\begin{equation}\label{e13}
\Gamma[f(t)]-\Gamma[\overline{f}(t)]\propto\left(e^{-\Gamma_L^{(s)}t}-
e^{-\Gamma_H^{(s)}t}\right)\sin\varrho_f\,\sin\gamma\,.
\end{equation}
Here $\varrho_f$ is a CP-conserving strong phase shift and $\gamma$ is the 
usual angle of the unitarity triangle. Because of the $\sin\varrho_f$ factor,
a non-trivial strong phase shift is essential in that case.
Consequently the CP-violating observables Eq.~(\ref{e13}) vanish within 
the factorization approximation predicting $\varrho_f\in\{0,\pi\}$. 
Since factorization may be a reasonable working assumption 
for the color-allowed modes $\stackrel{{\mbox{\tiny 
(---)}}}{B_s}\to D_s^\pm K^\mp$, the CP-violating effects in their
untagged data samples are expected to be very small. On the other hand,
the factorization hypothesis is very questionable for 
the color-suppressed decays $\stackrel{{\mbox{\tiny (---)}}}{B_s}\to
\stackrel{{\mbox{\tiny (---)}}}{D^0}\phi$ and sizable CP violation may 
show up in the corresponding untagged rates.\cite{dunietz}\, 

Concerning such CP-violating effects and the extraction of $\gamma$ from
untagged $B_s$ decays, the modes $\stackrel{{\mbox{\tiny (---)}}}{B_s}\to 
D_s^{\ast\pm} K^{\ast\mp}$ and $\stackrel{{\mbox{\tiny (---)}}}{B_s}\to
\stackrel{{\mbox{\tiny(---)}}}{D^{\ast0}}\phi$ are expected to be 
more promising than the transitions discussed above. As was
shown in Ref.~\cite{fd2}, the time-dependences of their untagged angular
distributions allow a clean extraction of $\gamma$ without any additional 
input. The final state configurations of these decays are not admixtures 
of CP eigenstates as in Section~\ref{admixtures}. They can instead be 
classified by their parity eigenvalues. A characteristic feature of the 
angular distributions is interference between parity-even and parity-odd 
configurations that may lead to potentially large CP-violating effects 
in the untagged data samples even when all strong phase shifts vanish. 
Therefore one expects even within the factorization approximation, 
which may apply to the color-allowed modes $\stackrel{{\mbox{\tiny 
(---)}}}{B_s}\to D_s^{\ast\pm} K^{\ast\mp}$, potentially large CP-violating
effects in the corresponding untagged data samples.\cite{fd2}\,
Since the soft photons in the decays $D_s^\ast\to D_s\gamma$, $D^{\ast0}
\to D^0\gamma$ are difficult to detect, higher resonances exhibiting
significant all-charged final states, e.g.\ $D_{s1}(2536)^+\to
D^{\ast+}K^0$, $D_1(2420)^0\to D^{\ast+}\pi^-$ with $D^{\ast+}\to 
D^0\pi^+$, may be more promising for certain detector configurations. 
A similar comment applies also to the mode $B_s\to D_s^{\ast+}D_s^{\ast-}$ 
discussed in Subsection~\ref{gold}.
\section{Conclusions}
Whereas $B_s\to\rho^0K_{\rm S}$ is expected to be the ``wrong'' way to
extract $\gamma$ because of hadronic uncertainties related to penguin 
contributions, there are other $B_s$ decays which should allow meaningful 
determinations of this CKM angle. Some of these strategies are even 
theoretically clean and suffer from no hadronic uncertainties. 

Within the Standard Model one expects very rapid $B_s^0-\overline{B_s^0}$
oscillations which may be too fast to be resolved with present vertex
technology. However, the corresponding $\Delta M_st$ terms cancel in 
untagged rates of $B_s$ decays that depend therefore only on the two 
exponents $e^{-\Gamma_L^{(s)}t}$ and $e^{-\Gamma_H^{(s)}t}$. If the width 
difference $\Delta\Gamma_s$ is sizable -- as is expected from theoretical 
analyses -- untagged $B_s$ decays may allow the determination both of the 
CKM angle $\gamma$ and of the Wolfenstein parameter $\eta$. Following these
lines one may furthermore obtain valuable insights into the mechanism of 
CP violation thereby getting indications for physics beyond the Standard 
Model. 

Obviously the feasibility of such untagged $B_s$ strategies to search for 
CP violation and to extract CKM phases depends crucially on a sizable 
width difference $\Delta\Gamma_s$. Moreover a lot of statistics is required 
so that hadron machines seem to be the natural 
place for such experiments. It is not yet clear whether the $B_s$ width 
difference will turn out to be large enough to make these measurements
possible. However, even if it should be too small, once 
$\Delta\Gamma_s\not=0$ has been found experimentally, the formulae 
developed in Refs.~\cite{fd1,fd2} have also to be used to determine CKM 
phases correctly from tagged $B_s$ decays. Certainly time will tell and 
hopefully an exciting future of CP violation in $B_s$ decays is ahead of
us. 
\section*{Acknowledgment}
I am grateful to Isi Dunietz for a collaboration on topics 
presented in this talk.

\end{document}